\documentclass[a4paper,12pt,oneside]{article}
\usepackage[koi8-r]{inputenc}
\usepackage[english,russian]{babel}
\usepackage{graphicx}
\usepackage{amsmath,amssymb}			

\begin{document}

\title{Anomalous eclipses of the young star RW~Aur~A}

\author{
S.~Lamzin$^1$ \footnote{E-mail: lamzin@sai.msu.ru}, D.~Cheryasov$^1$, G.~Chuntonov$^2$, \\
A.~Dodin$^1$,  K.~Grankin$^3$, K.~Malanchev$^1$, A.~Nadzhip$^1$, \\
B.~Safonov$^1$, D.~Shakhovskoy$^3$, V.~Shenavrin$^1$, A.~Tatarnikov$^1$, O.~Vozyakova$^1$ \\
}

\date{
$^1$ Sternberg Astronomical Institute, M.~V. Lomonosov Moscow State
University, Moscow, Russia;\\
$^2$  Special Astrophysical Observatory, Russian Academy of Sciences,\\ Nizhnii
Arkhyz, Russia \\
$^3$ Crimean Astrophysical Observatory, p/o Nauchny, Republic of Crimea \\
}

\maketitle


\section*{Abstract}
Results of $UBVRIJHKLM$ photometry, $VRI$ polarimetry and optical spectroscopy
of a young star RW~Aur~A obtained during 2010-11 and 2014-16 dimming events
are presented. During the second dimming the star decreased its brightness to
$\Delta V > 4.5$ mag, polarization of its light in I-band was up to 30\,\%, and 
color-magnitude diagramm was similar to that of UX~Ori type stars. We
conclude that the reason of both dimmings is an eclipses of the star by dust 
screen, but the size of the screen is much larger than in the case of UXORs. 

\medskip

\section*{Introduction}

  RW~Aur is a young binary \cite{Joy44} with current separation between
components $\approx 1.5^{\prime\prime}$.  The primary of the system RW~Aur~A
is a classical T Tauri star, i.e.  low mass pre-main-sequence star, which
accretes matter from a protoplanetary disk \cite{P01}.  \cite{C06} found a
spiral arm of molecular gas going out from the disk and concluded "that we
are witnessing tidal stripping of the primary disk by the recent fly-by of
RW~Aur~B".  Hydrodynamical simulations of \cite{D16} confirm this tidal
interaction hyphothesis.

   Recently RW~Aur~A has undergone two major dimming events with
unprecedented parameters. The first one has occured in 2010-11 ($\Delta t
\sim 150^d,$ $\Delta V \sim 2$ mag) and the second even deeper dimming was
from summer 2014 to summer 2016 \cite{R13,R16}. Non-trivial behavior of 
the star during these events was discussed in a number of papers -- see e.g. 
\cite{A15}, \cite{P15}, \cite{Sc15}, \cite{Sh15}, \cite{B16}, \cite{F16},
\cite{T16} and references therein. There are no doubts now that
both dimming events were due to eclipse of the star by dust screen, but the
nature of the screen is a matter of debates: a tidal arm between components
of the binary, dusty disk wind or/and warped inner disk regions.

   We present here preliminary results of our observations during both
dimming events and some general conclusions.

\section*{Observations and Results}

  Our observations of RW~Aur were carried out at Crimean Astrophysical
Observatory (CrAO, 1.25-m telescope), Special Astrophysical Observatory (SAO,
6-m telescope), and at two observatories of Lomonosov Moscow State
University: Caucasus Mountain Observatory (CMO, 2.5-m telescope) and Crimean
observatory (CO, 1.25-m telescope). Photometric and polarimetric data in
optical band were obtained at CrAO (2010-2011, non-resolved) and CMO
(2014-2016, resolved). Medium ($R\approx 15000)$ resolution spectra in
5850-6600 \AA\AA\, band were observed at SAO in November 2014 and December
2015. Infrared photometric data were obtained at CO (2010-2016, $JHKLM,$
non-resolved) and CMO (2015-2016, $JHK,$ resolved). Details of the
observations and data reduction will be presented in forthcoming paper.

\begin{figure}[]
 \begin{center}
  \resizebox{14.0cm}{!}{\includegraphics{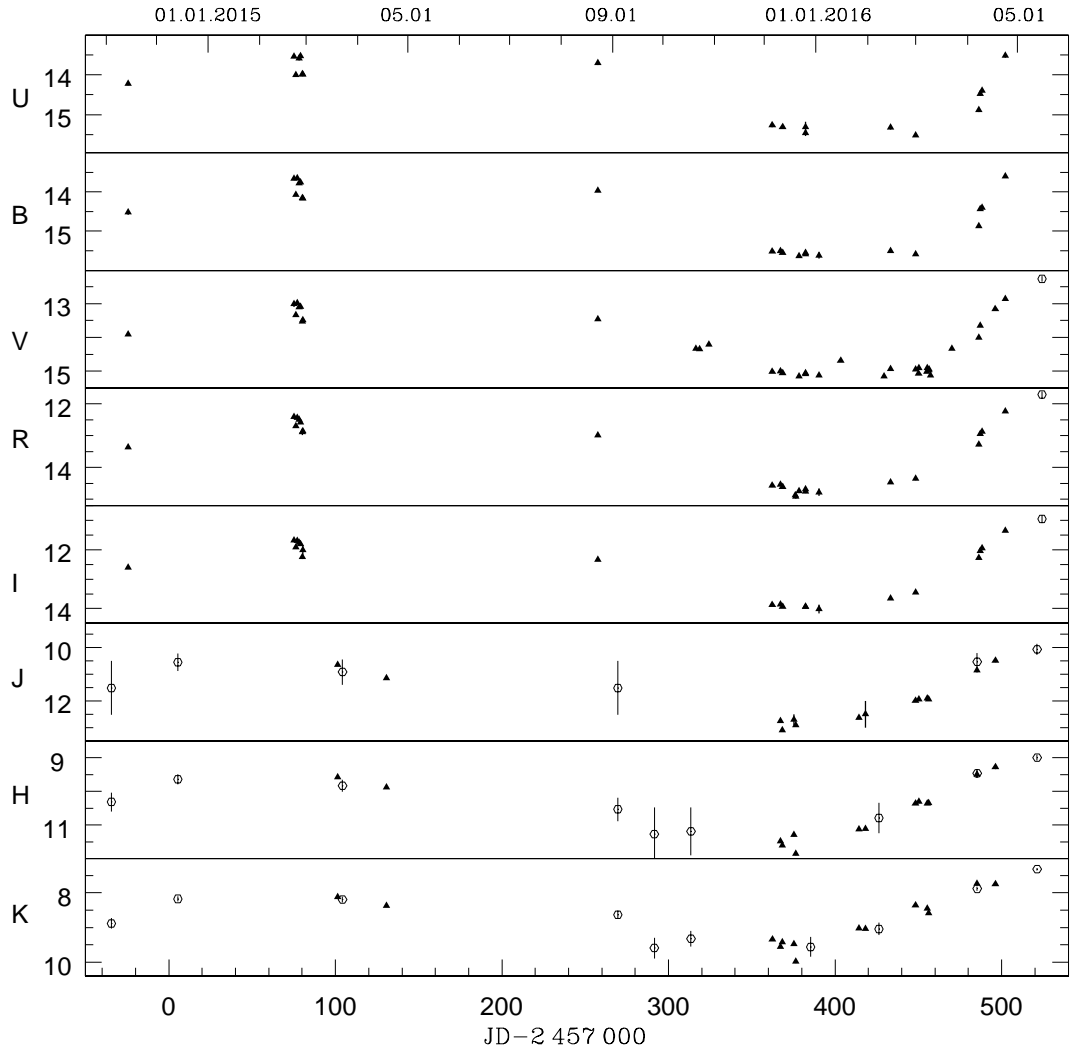} }
  \caption{ $UBVR_c$I$_cJHK$ light curves of RW~Aur~A
during the second dimming event. Filled triangles -- resolved measurements of the
binary, open circlues -- non-resolved data corrected for avarage contribution of
RW~Aur~B.
}
  \label{fig1}
  \end{center}
\end{figure}

 It can be seen from Figure~\ref{fig1} that 2014-16 dimming of RW~Aur~A
occured in all spectral bands from $U$ to $K.$ The larger wavelength the
smaller an amplitude $\Delta m$ of the dimming: if values $V=10.5^m$
\cite{H94}, $K=7.2^m$ \cite{Sh15} are adopted as an average values before 2010, then
$\Delta V \approx 4.6^m,$ $\Delta K \approx 2.8^m.$ Minimal brightness was
reached near the end of 2015 and then there was a period of almost constant
brightness (plateau), which was shorter in NIR band than in optical one.

\begin{figure}[]
 \begin{center}
  \resizebox{14.0cm}{!}{\includegraphics{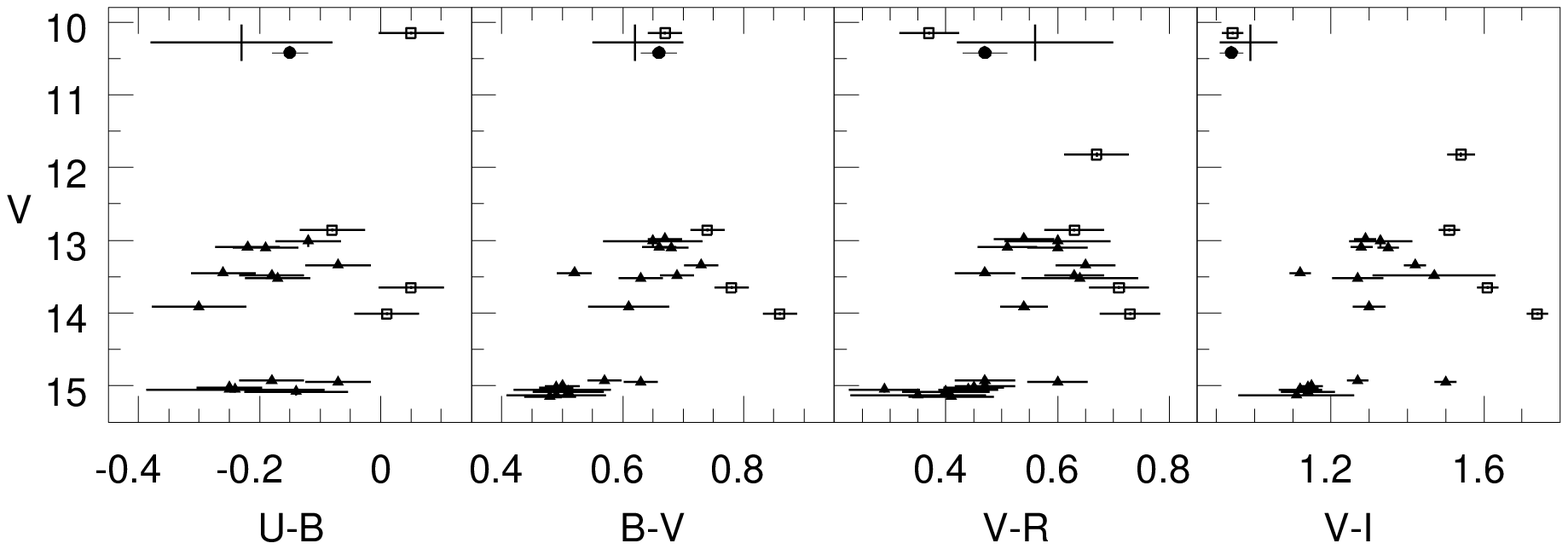} }
  \caption{ Color-magnitude diagramm of RW~Aur~A
during 2014-2016 dimming. Filled triangles refer to ingress and plateau 
phases, open circlues -- to egress phase. Cross and solid circle
represent data before 2010: non-resolved photometry of \cite{H94} and
resolved photometry of \cite{WG01}, respectively.
}
  \label{fig2}
  \end{center}
\end{figure}

 As follows from Figure~\ref{fig2} a color-magnitude diagramm of the star
during 2014-16 dimming is similar to that of so called UXORs \cite{G94}:
initially RW~Aur~A becomes redder when it fades, but at $V \ge 13.5^m$ its
colors become bluer when the star dims further. Note that colors during
engress are redder than during ingress.

\begin{figure}[]
 \begin{center}
  \resizebox{14.0cm}{!}{\includegraphics{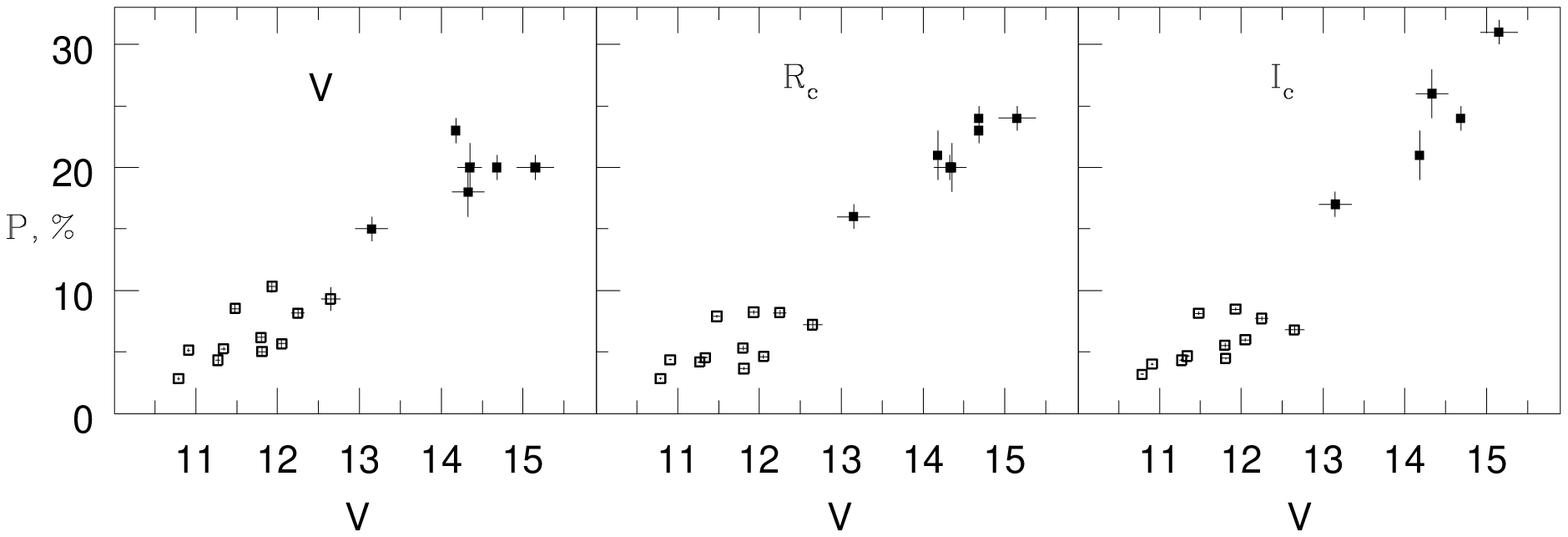} }
  \caption{ Polarization in $V,$ $R_c$ and $I_c$ bands
vs $V$-magnitude. Filled squares -- resolved polarimetry of RW~Aur~A during the
second dimming, open squares -- unresolved polarimetry of both components 
during the first dimming with subtracted contribution of RW~Aur~B.
}
  \label{fig3}
  \end{center}
\end{figure}

  Optical $(UBVRI)$ radiation of the star during 2010-11 and 2014-16
dimmings was strongly linearly polarized as in the case of UXORs
\cite{G94}. The degree of polarization increases when the star fades and
rises with wavelength: at the plateau phase it reached 30\% in $I$ band (see
Figure~\ref{fig3}), but was $<1\%$ before 2010 \cite{V05}. Position angle
of polarization during both dimmings was parallel to that of major axis of
RW~Aur~A's disk, while radiation transfer modelling of protoplanetary disks
predicts that they should be perpendicular \cite{KW14}.
  
  \cite{Sh15} observed increase of the 3-5 $\mu$m flux of RW~Aur at the
beginning of 2014-2016 dimming at wavelength $\lambda \leqslant 2.2$ $\mu$m. 
We report that brightness of the star returned back to pre-dimming level in
both these spectral bands by the end of summer 2016.

  We found from our spectra that equivalent width of iron emission lines
increased only in a few times when observed continuum flux decreased dozens
of times.  Note also that profiles of these lines during ingress and plateau
phases of 2014-16 dimming were asymmetric: their redward part was stronger
than blue one.

  Our results confirm that both dimming events were due to eclipse of
RW~Aur~A by dust cloud, grains of which produce selective absorption.  In
some respects the behavior of the star reminds that of UXORs, but duration
and amplitude of 2014-16 eclipse are much larger, linear polarization of
light is much stronger and has another orientation. In our case dust screen
eclipses not only the star, but also significant part of scattering disc and
a region where iron emission lines are originated. We don't know the nature
of the dust screen (a tidal arm, dusty disk wind etc.), but its origin
undoubtly connected with tidal disturbance of RW~Aur~A disk by recent fly-by
of the companion.

\medskip

\it{Acknowlegement} This study was partially supported by grants
RFBR-16-02-00140, RFBR-16-32-60065, NSh-6579.2016.2 and Lomonosov Moscow
State University Program of Development.

\end{document}